\documentclass[fleqn,twoside]{article}
\usepackage{espcrc2,amssymb}
\usepackage{graphicx,epsfig,color}

\newcommand{\AmS}{{\protect\the\textfont2A\kern-.1667em\lower.5ex\hbox{M}\kern-.125emS}}
\title{10-100 TeV cosmic ray anisotropy measured at Baksan EAS "Carpet" array}
\author{V.V. Alekseenko\thanks{e-mail: vicalek@rambler.ru},
 A.B. Cherniaev, D.D. Djappuev, N.F.Klimenko,
  A.U. Kudjaev, O.I. Michailova, Yu.V. Stenkin,
V.I.Stepanov, V.I. Volchenko
\address{Institute for Nuclear Research of
Russian Academy of Sciences\\
        117312, Moscow, RUSSIA}}%
\begin{document}
\begin{abstract}
Preliminary results of one year anisotropy measurement in the
energy range $10^{13} -10^{14}$ eV as a function of energy are
presented. The results are compared for two methods of data
analysis: the standard one with meteo correction approach in use
and another one so-called "East minus West" method.  Amplitudes
and phases of anisotropy for three median energies  E = 25 TeV,
E = 75 TeV and E = 120 TeV are reported. Brief consideration of
amplitude-phase dependence of anisotropy on energy is expounded.
\end{abstract}
\maketitle
\section{Introduction}

Experimental results on anisotropy of primary cosmic ray flux give
important information for further theoretical study of the origin
and propagation problems in cosmic ray physics. Together with
numerical characteristics of anisotropy - amplitude and phase -
the energy dependence of these characteristic gives additional
information for better understanding of the above mentioned
problems. Energy range of primary cosmic rays under anisotropy
investigation is extremely wide - from $\sim$ 100 GeV up to
$10^{5}$ TeV and higher.  At 1 - 10 TeV range the measurements are
carried out with detectors of Cerenkov emission of Extensive Air
Showers or underground muon telescopes (Nagoya, Baksan Underground
Scintillation Telescope; MACRO and others). Then, at 100-10000 TeV
range, the EAS-TOP and CASCADE results are well known. And finally
for $10^{17}$ eV and above anisotropy is studied with giant EAS
installations such as Yakutsk, AGASA and Auger. One can find
review of anisotropy results in $\cite{ghi2}$ and bibliography in
$\cite{ghi2,ghi,agl}$. Some results of cosmic ray anisotropy
studies with a Small Air Shower (SAS) arrays (E $\sim$ 10 TeV) in
Northern hemisphere were published at seventieth-eightieth
$\cite{sak,gom,nag,ale1,ale2}$. Amplitude and phase of anisotropy
were derived from Fourier analysis of counting rate along the
Right Ascension coordinate, without record of SAS arrival
direction. The main results of those studies were:  1).
Compton-Getting effect due to revolution of the Earth around the
Sun in accordance with theoretical prediction $\cite{com}$ was
observed, 2). Sidereal anisotropy with amplitude $\sim 0.06\%$ and
phase $\sim$ 1 hour RA was measured with high statistical
accuracy. When there are numerous results for E $<$ 10 TeV and E
$>$ 100 TeV, the 10-100 TeV range is up to now slightly studied.
Lack of experimental results at this energy range gives no
possibility to arrive at clear conclusion about dependence of
amplitude and phase of anisotropy on energy. At 2007, after some
modernization of Baksan EAS installation, we resumed the
measurement of anisotropy of SAS in the range 10-100 TeV with
registration of arrival directions on celestial sphere. Below we
report result of one year registration.
\section{Experimental set-up and Energy Response}

Central part of Baksan SAS array - "Carpet" - consists of 400
liquid scintillation detectors, arranged in horizontal continuous
square geometry - Fig.1. Dimension of the   detector is 70x70x30
cm$^{3}$. There are four outside huts (OH) on square diagonals at
distance 30 m from the center of Carpet. Each hut contains  18
(3x6)detectors. The Carpet is divided into 25 square modules, each
module = 16 detectors. Twenty nine (25 modules + 4 OH) sum anode
pulses (each pulse is a sum of anode pulses of individual
detectors of module) are registered with ADC, TDC and with
multiple logic unit designed to produce  the  SAS triggers.
Individual (each of 29) logic pulse is generated if corresponding
sum anode pulse exceeds level corresponding to energy deposition
of 0.5 relativistic particle. Energy deposition corresponding to 1
r.p. is equal 50 MeV.  Multiplicity of individual logic signals
together with anode amplitudes are the measure of power of SAS,
following the primary particle energy.Arrival direction of SAS is
calculated using the relative times from TDC channels. In the
present analysis we have used three sorts of triggers:4-fold
coincidence of corner modules 4CM, 4-fold coincidence of outside
huts 4OH, 8-fold coincidence 4CM+4OH. Counting rate of
coincidences: F(4CM)=1.39 Hz, F(4OH)=1.50 Hz, F(4CM+4OH)=0.72 Hz.
Angular resolution: R(4CM)=$10^o$, R(4OH)=$3.5^o$,
R(4CM+4OH)=$3.0^o$.One year data set was processed, only full days
were included in analysis,useful time after "noise" cuts and
elimination of "bad" days = $75\%$.Energy response for three sorts
of trigger was evaluated by Monte Carlo simulations, made on the
basis of CORSIKA codes (v. 6012, HDPM and Gheisha models).
Vertical proton flux (with differential spectrum $\gamma$ = - 2.7)
uniformly distributed over the circle area with r = 60 m was used
in simulation. The roof of thickness 34 $g/cm^{2}$above the Carpet
was taken into account. Evaluated median energies for three sorts
of trigger are: E (4CM) =  25 TeV, E (4OH) = 75 TeV,  E (4CM+4OH)
= 120 TeV.
\begin{figure}[tbh]
\vspace{-1.pc} \epsfig{height=5.5cm,file=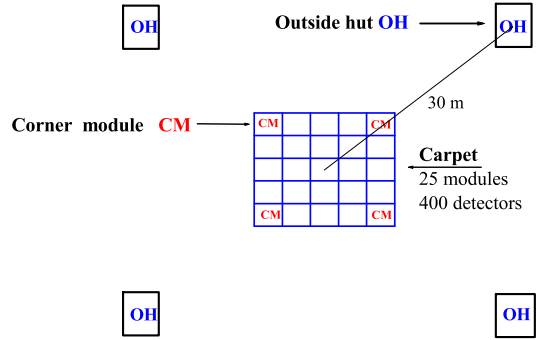} \caption{Plan of
the Baksan  SAS array.} \label{fig1}
 \end{figure}
%
\section{Processing and analysis of experimental information.}

Fourier analysis of resulting day waves in solar, sidereal and
antisidereal time was done to find amplitudes and phases of the
1st and 2nd harmonics. Resulting day waves were obtained by two
methods with posterior comparison of methods' efficiency: a)
traditional approach with meteo correction, b) differential method
"East minus West".  The following considerations were used as
efficiency criteria: 1) Phase of the first solar harmonic must be
close to 6 hours of Local Solar Time (Compton-Getting effect), and
amplitude to be close to 0.034  for our geographic latitude. 2)
Absence of statistically significant antisidereal wave. It means
the absence of annual modulation of solar day wave. Otherwise the
false sidereal wave inevitably arises as one of two side
frequencies - sidereal and antisidereal.
  \subsection{Traditional approach}

Counting rate in cosmic ray variation experiments depends
essentially on variations of meteorological factors - atmosphere
pressure and temperature. Hence, one need to exclude meteo induced
variations to get pure initial variations of counting rate.
Meteorological coefficients for three sorts of trigger were
derived from multiple regression analysis. Amplitudes and phases
of 1st harmonics after meteo correction are presented in Table 1.
All 2nd harmonic amplitudes are inside statistics. It is necessary
to note that comparatively large statistical error arises from
short enough data set - only one year of registration. And
nevertheless, phases of solar waves are in good coincidence with
expected one from Compton-Getting effect - all three are close to
6 hour of Local Solar Time.  At the same time extremely large
($>5\sigma$) amplitude of 75 TeV antisidereal wave gives rise to
assume that meteo correction procedure by itself doesn't guarantee
total expulsion of extraneous variations. This fact arouses doubts
about sidereal time result.
\begin{table*}[htbp!]
\caption{Amplitudes and phases of 1st harmonic after meteo
correction.The amplitude errors shown in Table1 and Table2 are statistical ones.} \label{table:1}
\newcommand{\m}{\hphantom{$-$}}
\newcommand{\cc}[1]{\multicolumn{1}{c}{#1}}
\renewcommand{\tabcolsep}{1.75pc} 
\renewcommand{\arraystretch}{1.1} 
\begin{tabular}{@{}lllll}
\hline
        wave                 & \m Solar  &\m Sidereal & \m Antisidereal \\
1st harmonic\\
\hline
25 TeV   amplitude, $\%$           & \m $0.025\pm0.025$ &\m $0.065\pm0.025$ &\m $0.064\pm0.025$  \\
phase, h                           & \m $6.4\pm2.3$  & \m $23.5\pm1.4$ &\m$19.1\pm1.4$ \\
75 TeV  amplitude, $\%$            & \m $0.050\pm0.024$ &\m $0.088\pm0.024$ &\m $0.122\pm0.024$   \\
phase, h          & \m $7.5\pm1.7$ &\m $19.6\pm1.0$ &\m $1.2\pm0.7$     \\
120 TeV   amplitude, $\%$          & \m $0.047\pm0.035$ &\m $0.059\pm0.035$ &\m $0.087\pm0.035$   \\
phase, h          & \m $6.8\pm2.2$ &\m $21.1\pm1.7$ &\m $22.4\pm1.5$     \\
\hline
\end{tabular}\\[2pt]
\end{table*}
\subsection{East minus West method.}

When studying anisotropy we come across not only with variations
induced by meteo factors but also with apparatus instabilities. To
remove them totally or simply to take into account their effect
is not so easy task. Wit way to solve the problem is the 'East minus
West" method mentioned for the first time in $\cite{nag2}$. The
method is based on the assumption that as meteo factors so
apparatus instabilities produce equal variations in counting rates
of showers arriving from East and West directions. Hence the
difference between East-ward and West-ward counting rates can
eliminate the uncontrolled and spurious variations. Sequence of
differences of counting rates from two directions calculated
during the day for each fixed time interval (20-min in our case)
results in differential day wave.Then integration of resulting
(sum of all days) differential wave results in primordial day wave
characterized by amplitude and phase. Amplitudes and phases of 1st
harmonics resulting from East-West method are presented in Table
2.
\begin{table*}[htb!]
\caption{Amplitudes and phases of 1st harmonic after East minus
West method.}
\label{table:2}
\newcommand{\m}{\hphantom{$-$}}
\newcommand{\cc}[1]{\multicolumn{1}{c}{#1}}
\renewcommand{\tabcolsep}{1.75pc} 
\renewcommand{\arraystretch}{1.1} 
\begin{tabular}{@{}lllll}
\hline
        wave                 & \m Solar  &\m Sidereal & \m Antisidereal \\
1st harmonic\\
\hline
25 TeV   amplitude, $\%$          & \m $0.046\pm0.025$ &\m $0.060\pm0.025$ &\m $0.045\pm0.025$  \\
phase, h                         & \m $6.0\pm1.9$  & \m $21.5\pm1.5$ &\m$7.9\pm1.4$ \\
75 TeV  amplitude, $\%$          & \m $0.063\pm0.024$ &\m $0.043\pm0.024$ &\m $0.068\pm0.024$   \\
phase, h                         & \m $6.8\pm1.4$ &\m $22.0\pm1.9$ &\m $5.8\pm0.7$     \\
120 TeV   amplitude, $\%$        & \m $0.032\pm0.035$ &\m $0.026\pm0.035$ &\m $0.011\pm0.035$   \\
phase, h                         & \m $8.3 $ &\m 0.4 &\m 20.7   \\
\hline
\end{tabular}\\[2pt]
\end{table*}
Comments to Table 2: 1) Antisidereal waves are inside of
statistical uncertainties in contrast to meteo method. 2) Phases
of solar waves are in good agreement with Compton-Getting effect.
3) There is also a hint that sidereal amplitude decreases with
increase of energy - on Fig.2 points of this experiment are
connected with dot line.
\begin{figure*}[tbh!]
\vspace{-1.pc}
\epsfig{height=9.cm,file=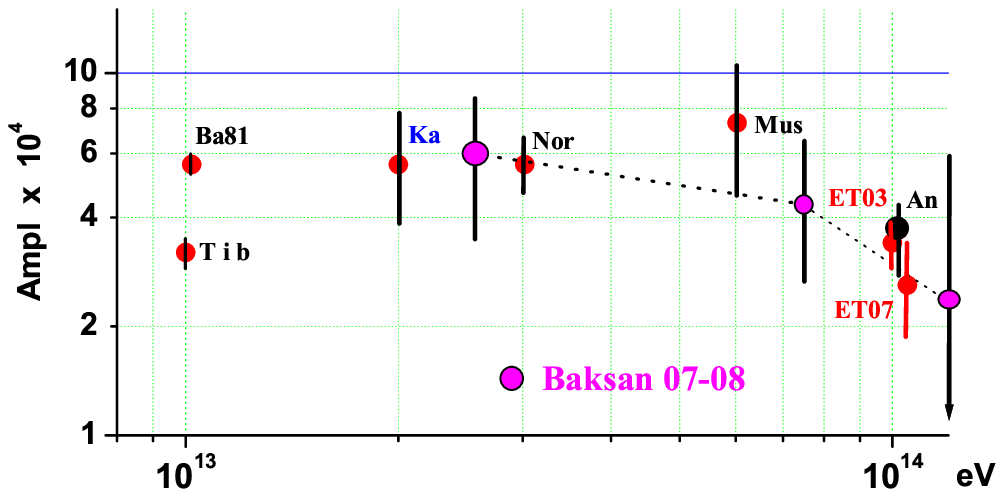}
\caption{Amplitude of 1st harmonic of sidereal
anisotropy for 10 -100 TeV energy range.}%
Ba81 - \cite{ale1,ale2}, Tib - \cite{yi}, Ka - \cite{oy},%
 Nor - \cite{sak,nag},
 Mus - \cite{gom}, ET03 - \cite{eas}, ET07 - \cite{ghi}, An - \cite{koz}
\label{fig2}
\end{figure*}
%
Earlier EAS-TOP team also reported comparatively small amplitude
$\cite{eas}$ for E = 100 TeV. One year later Andyrchi result
$\cite{koz}$ for the same energy was presented being in excellent
agreement with EAS-TOP one. It seems worthwhile to do some remarks
on possible energy dependence of anisotropy.  At 30th ICRC the
EAS-TOP collaboration presented $\cite{ghi}$ results of
reprocessed 8-year data set using E-W method - point ET07 at
Fig.2. We can notice that this result together with previous ones
for E = 100 TeV really indicate decreasing of amplitude for energy
close to 100 TeV.  And what is more interesting, with further
increasing of energy the amplitude  again increases up to 0.064 \%
at E = 400 TeV with drastic change in phase $\cite{ghi,agl}$ to
13.6 hour RA from 0.4 hour RA for 100 TeV. Apropos, such shift of
phase was already pronounced in $\cite{eas}$ for energy ~ 300 TeV
and higher but with less significance. Phase of max counting rate
0.4 hour RA for low energy corresponds to direction perpendicular
to local Orion arm (in out of arm direction) and parallel to
Galactic plane whereas 13.6 hour RA points to north Galactic Pole
inside the local Galactic arm. Hence, it is possible to suppose
that something changes in mechanism producing an anisotropy effect
in primary cosmic ray flux at energy range 100 - 300 TeV.
Obviously, to arrive at more definite conclusion we need acquire
more data.
\section{Conclusion}

Posterior analysis demonstrates that "East-West" method yields
more reliable result on sidereal anisotropy in comparison with
standard meteo correction method. Further accumulation of data is
necessary to understand better the evolution of cosmic ray
anisotropy in the energy range 10 - 100 TeV. Together with results
from high energy range it will bring some new knowledge about
origin and propagation of cosmic rays.

The work was supported in part by the RFBR grant N 06-02-16355, by
the Government Contract N 02.445.11.7070 and by the RAS Basic
Research Program "Neutrino Physics".

\end{document}